%% file: main.tex
\documentclass[%
 reprint,
 amsmath,amssymb,
 aps, showkeys, prmaterials
]{revtex4-2}

\usepackage{dcolumn}
\usepackage{bm}

\usepackage[utf8]{inputenc}
\usepackage{graphicx}
\usepackage[labelformat=simple]{subcaption}
\usepackage{miller}
\usepackage{siunitx}
\usepackage[british]{babel}
\usepackage{ifthen}
\usepackage{booktabs}
\usepackage{xfrac} 
\usepackage{paralist}
\usepackage{mhchem}
\DeclareSIUnit{\atperc}{at.\%}

\input{commands}

\begin{document}

\preprint{APS/123-QED}

\title{Interaction of monoclinic \zro{} grain boundaries with oxygen vacancies, Sn and Nb\,---\,implications for the corrosion of Zr alloy fuel cladding}

\author{M.\,S.\,Yankova}
 \email{maria.yankova@manchester.ac.uk}
\author{C.\,P.\,Race}
\affiliation{Department of Materials, University of Manchester}

\date{\today}

\begin{abstract}
 We used density-functional-theory simulations to examine the structural and electronic properties of the \gbsymb{} grain boundary in monoclinic \zro{}, which is a very low-energy (\jpermsq{0.06}) twin boundary present in experimental oxide texture maps, with suggested special properties. This equilibrium structure was compared with a metastable structure (with a boundary energy of \jpermsq{0.32}), which was considered to be representative of a more general oxide boundary. The interaction of oxygen vacancies, substitutional Sn and Nb defects (substituting host Zr sites) with both structures\,---\,and their effect on the boundary properties\,---\,were examined. We found that the equilibrium structure energetically favours \vo{2+} and \nb{Zr}{2-}, whereas the metastable structure favours \vo{2+} and \sn{Zr}{2-}. Tin was further found to bind strongly with oxygen vacancies in both structures, and introduce gap states in the band gap of their electronic structure. \sn{Zr}{2-} was, however, found to increase the segregation preference of \vo{2+} for the metastable structure, which might contribute to increased oxygen and electron transport down this interface, and therefore other more general boundaries, compared to the equilibrium structure of the studied monoclinic twin boundary.
\end{abstract}

    
\maketitle

\section{\label{sec:level1}Introduction}

Improving the understanding of oxygen transport in zirconium oxide, and the interaction of oxygen with microstructural planar defects such as grain boundaries (GBs), and dopant point defects, is a key factor in achieving a more mechanistic knowledge and better control of the corrosion of zirconium alloys, of which the cladding tubes in light water nuclear reactors are constructed. Corrosion of zirconium alloys (and the associated hydrogen pick-up), is one of the major degradation mechanisms that limits the lifetime of the cladding, and hinders the possibility of extending the fuel burn-up. For these reasons, corrosion has been a focus of research interest since the first Zr alloy was introduced in the 1960s. However, a complex combination of a very fine oxide grain size, multiple phases, alloying elements, irradiation, stress and a harsh environment, has made it difficult to unravel the governing mechanisms of corrosion; various hypotheses exist.

Zirconium is a very reactive metal, which forms a thin oxide layer in any oxygen-containing environment. The oxide layer grows continuously inwards due to diffusion of oxygen from the oxide-water interface to the metal-oxide interface \cite{Woolsey1981}. Typical corrosion kinetics of Zr alloys exhibit a cyclic behaviour, where the formation of a protective oxide layer begins with an initial ‘pre-transition’ stage, which follows a close-to-cubic or parabolic law\,---\,depending on the alloy\,---\,up to an oxide thickness of about 2 microns \cite{Cox2005}. 
During this stage, a gradient of oxygen concentration develops from the metal towards the water environment from the solid solution limit of \SI{29}{\atperc} to the stoichiometric oxide of \SI{66.67}{\atperc} \cite{Yilmazbayhan2006}. As a consequence of the oxygen sub-stochiometry, there is an observed phase change from Zr metal to Zr-O solid solution, to a sub-oxide ZrO phase, to a mixture of stable monoclinic and small amounts of metastable tetragonal or cubic equiaxed \zro{} grains in the range \SIrange{10}{30}{\nano\meter} \cite{Godlewski1991, Barberis1995, Motta2005, Ni2011, Gabory2015}. The metastable grains are stabilised by a combination of factors, including small grain size, stress and dopants \cite{Bouvier2002, Yilmazbayhan2004, Qin2007, Wei2013, Garner2015}. The Pilling-Bedworth ratio for the zirconium metal to oxide transformation is 1.56, which is equivalent to a nearly 60\% volume increase. The zirconia film is therefore under high planar compressive stress. A competition between epitaxial strain and growth stresses forms a microstructure consisting of columnar grains up to about \SI{200}{\nano\meter} long \cite{yankova2023}.  A strong preferential texture then develops in the oxide with varying degree of strength, depending on the alloy and/or amount of irradiation \cite{Yilmazbayhan2006}. 
We have shown in previous work \cite{yankova2023} that in Zr alloys with split-basal texture, the growth stress is the main driving force for the growth of specific grain orientations from a strong one-component epitaxial texture by minimising areal footprint and stiffness. As a result, the protectiveness of the oxide layer is shown to depend on the formation of coarser, more coherent microstructures with wider columnar grains which are thought to have a higher fraction of protective, low energy grain boundaries. In particular, a high fraction of monoclinic twin boundaries have been observed, which are proposed to be difficult migration paths due to their high coherency and low grain boundary energy \cite{Gertsman1997, Gertsman1999, Garner2014, Garner2015}, though there are no previous studies estimating the twin GB energy and properties in \zro{m-}. Twin GBs can be formed during the nucleation of new monoclinic grains. Alternatively, transformation twins have been observed as a result of the tetragonal stress-stabilised grains transforming to monoclinic due to the stress gradient in the oxide layer \cite{Qin2007}. At a critical oxide thickness of about \SI{2}{\micro\meter}, an acceleration in the corrosion kinetics is observed, referred to as ‘transition', which is associated with a breakdown of the columnar grain growth and a large number of defects forming at the oxide front. It is followed by a recovery in the protectiveness of the oxide layer and the beginning of another cycle. The process is repeated until the oxidation reaches the final stage of the corrosion kinetics, termed ‘breakaway’, which is characterised by approximately linear growth.

Although the exact species transport mechanisms are unknown, oxygen, hydrogen and electron transport have been identified as the most important in the oxidation process, since zirconium was shown to be immobile under standard reactor operating conditions \cite{Whitton1968, Kilo2003}. From an electrochemical point of view, a cathodic reaction occurs at the oxide-water interface: water is dissociated, producing oxygen ions and protons, which both diffuse inwards independently, either as point defects in the bulk lattice or down extended defects, such as GBs, dislocations and regions of porosity \cite{Brossmann1999, Foster2002a, Garner2018}. An alternative hypothesis suggests O and H travel together as an \ce{OH$^-$} group. Although OH$^-$ was found to be immobile in single crystal tetragonal grains, some studies \cite{Lindgren2014} have suggested \ce{OH^-} transport might be relevant near interfaces. An anodic reaction occurs at the metal-oxide interface: zirconium atoms combine with the oxygen ions to form zirconium oxide and also create oxygen vacancies and free electrons, which diffuse outwards towards the oxide-water interface.
If we assume charge neutrality within the oxide layer, and since electron transport is slow in a wide-band-gap insulator such as \zro{}, electronic conduction is expected to be the oxidation-rate determining process\,---\,this hypothesis is supported by the small negative potential always measured on the Zr metal \cite{Yilmazbayhan2004}. Another important role of the electron conductivity is the recombination of the electrons and protons somewhere within the oxide layer to form atomic hydrogen. Based on how close or far from the metal-oxide interface that happens, a larger or smaller fraction of the hydrogen diffuses into the Zr metal (a phenomenon known as hydrogen pick-up (HPU)) and precipitates as brittle hydrides, which can crack under stress and lead to failure \cite{Efsing2000}.  
The main suggested mechanism for electron transport is via additional electronic states in the band gap of \zro{}, which can be introduced by defects and dopants, and which effectively reduce the band gap. Nishizaki et al.\,\cite{Nishizaki2002} used molecular orbital modelling to show that H introduces a new state below the conduction band. Alternatively, Cox et al.\,\cite{Cox2005} proposed electron conductivity through change of the valence state of the Zr vacancy, \ce{V^{4+}_{Zr}}, to lower valence states.

In a nano-crystalline material such as the Zr oxide layer, we would expect grain boundaries to be the dominant diffusion paths for species. Brossmann et al.\,\cite{Brossmann1999} used secondary ion mass spectroscopy to measure the oxygen diffusion through a nano-crystalline monoclinic \zro{} layer. They found that grain boundary diffusion was \numrange{e3}{e4} times faster compared to bulk diffusion for a grain size range of \SIrange{70}{300}{\nano\meter} within a temperature range of \SIrange{450}{950}{\celsius}, and assuming a grain boundary width of \nm{0.5} based on TEM observations \cite{Nitsche1996}. There is also evidence for deuterium enrichment at oxide grain boundaries using atom probe tomography, which suggests H diffusion is also preferable down boundaries \cite{Sundell2015a}. An et al.\,\cite{An2013} used aberration-corrected TEM combined with a hybrid Monte Carlo -- molecular dynamics model to show significant oxygen vacancy segregation to a $\Sigma$13 \ce(510)/\ce[001] symmetric tilt grain-boundary in yttria-stabilized zirconia (YSZ). Another first-principles study \cite{Marinopoulos2014} of a $\Sigma$5 \ce(310) tilt GB in YSZ found favourable segregation of neutral and negatively charged hydrogen. Oxygen transport through other metal oxides is also relevant. McKenna and Shluger \cite{McKenna2009a} reported a GB energy of \jpermsq{0.6} for the \hkl(101) twin GB in monoclinic \ce{HfO_2} and studied the interaction of oxygen vacancies with the GB in DFT. They found that both neutral and positive vacancies segregate favourably to the GB with segregation energies of --0.52 and \evolt{-0.8}, respectively. Additionally, positive vacancies were found to have high mobility parallel to the boundary but much lower mobility normal to the boundary. High GB concentrations of positive vacancies (\SI{2.5e14}{\per\square\centi\metre}) were found as possible percolation paths for electron tunnelling. The reported segregation energies of neutral O vacancies in other oxides include values between \evolt{-0.77} and \evolt{0.57} in a study of a \hkl(310) GB in \ce{SrTiO_2} \cite{Imaeda2008}, and between \evolt{-0.93} and \evolt{0.76} for a \ce(310)\ce[001] \degs{36.8} tilt GB in \ce{MgO} \cite{McKenna2009b}.

Two alloying elements that have opposing effects on the corrosion behaviour are tin and niobium. Tin was initially added to Zr alloys to improve the mechanical properties through solid solution strengthening and to oppose the detrimental effects of nitrogen and carbon. However, it was later found to have an adverse effect on oxidation, which led to the development of low-Sn, and Sn-free alloys \cite{Cox2005, Motta2015}. 
Studies have shown an improvement in oxidation behaviour with decreasing Sn content in autoclave \cite{Berry1961, Garde1994,Takeda2000, Wei2013}, and an even greater effect in reactor tests \cite{Garde1994}. In Zr-Sn binary alloys and alloys of the Zircaloy family, the observed effect is reduction in weight gain and corrosion rate \cite{Chirigos1952, Thomas1955, Berry1961, Garde1994,Takeda2000}, whereas in Zr-Sn-Nb alloys there is a delay in transition but no change in pre-transition corrosion kinetics \cite{Wei2013}. Therefore, an interaction between Sn and Nb is thought to exist, but its exact mechanism remains unknown. Moreover, tin in lower than 4+ charge states is thought to be stabilising the tetragonal phase---Sn is charge compensated by positive oxygen vacancies, and a high concentration of oxygen vacancies is a known mechanism for stabilising the tetragonal phase.
X-ray absorption near-edge spectroscopy (XANES) \cite{Hulme2016} and M\"ossbauer spectroscopy \cite{Pecheur2002} studies have shown that the average oxidation state of Sn in both Zircaloy-4 and ZIRLO was between 2+ and 4+, and there was metallic Sn and \ce{Sn^2+} present near the metal-oxide interface. Wei et al.\,\cite{Wei2013} demonstrated that with increasing tin content, the tetragonal phase fraction increases, followed by an increase in the tetragonal-to-monoclinic phase transformation, and therefore reduction in the protectiveness of the oxide layer. Similarly, dopants with valence state lower than that of \zr{4+}, such as \ce{Y^{3+}}, have been used to stabilise the tetragonal and cubic phases of \zro{} in its various other applications, including thermal barrier coatings \cite{Cao2002}, solid oxide fuel cell (SOFC) electrolytes \cite{Laguna-Bercero2012} and heterogeneous catalysis \cite{Paulidou2005}. Sn sits in solid solution in $\alpha$-Zr, however, there is some contradiction on how it is distributed in the oxide layer. Takeda et al.\,\cite{Takeda2000} showed Sn segregation to the \ce{ZrO} and \zro{} grain boundaries, as well as Sn localisation near the metal-oxide interface. Sundell et al.\,\cite{Sundell2015}, in contrast, used atom-probe tomography to show preference for metallic Sn particles and some clustering near oxide GBs but claimed no segregation. Bell et al.\,\cite{Bell2018} hypothesised that the reason for the lack of an increase in the corrosion rate, which one might expect if the oxygen vacancy concentration is increased in the presence of aliovalent ions such as \ce{Sn^{2+}}, is due to the strong binding of \vosn{2+}{2-}{0} clusters. This, however, does not explain the increase in the corrosion rate in alloys such as Zircaloy-4, which contain Sn but not Nb. 

On the other hand, tin-free Zr-Nb alloys exhibit lower corrosion rate compared to Zr-Sn-Nb and a further delayed transition \cite{Couet2015}. Additionally, Nb has a beneficial effect on the hydrogen pick-up \cite{Ramasubramanian2002, Bossis2006, Couet2013, Couet2015}, and thus most current reactor alloys contain Nb. Earlier theories suggested that Nb mainly exists in the 5+ charge state in the outermost oxide, therefore acting as an n-type dopant, donating electrons that combine with protons close to the water-oxide interface, and preventing H from diffusing through to the metal \cite{Ramasubramanian2002, Bossis2006}. This hypothesis was challenged by several XANES studies, which found Nb in various states from metallic Nb to oxidation states between 2+ and 5+ \cite{Froideval2008, Sakamoto2015, Couet2014, Couet2015}. Couet et al.\,\cite{Couet2015} developed a space-charge model, according to which the negative charge of the lower state Nb defects compensates the positive charge that builds up in the oxide layer due to an imbalance between the transport of vacancies and that of electrons, and thus reduces the hydrogen pick-up fraction (HPUF). Bell et al.\,\cite{Bell2017} used DFT studies to suggest that the low Nb oxidation states only exist in the tetragonal phase and form defect clusters with \vo{}. 
The corrosion behaviour of Zr alloys has improved dramatically through multiple alloy design iterations. However, a mechanistic understanding of the effect of alloying elements on the oxidation process is still needed, and would allow for any further improvements to be more economical.

There are a number of density functional theory (DFT) studies that have created a good basis of understanding of the defect chemistry in the bulk phases of \zro{} \cite{Foster2002a, Bell2015, Youssef2016, Bell2017, Bell2018, Yang2018}. However, the studies on grain boundaries are mostly limited to YSZ and cubic \zro{} \cite{An2013, Marinopoulos2014, Frechero2015}, as these phases are of interest in various other applications, and also due to the limitations of the first-principles methods, such as the requirement for periodic supercells and the limited system size. However, the majority of the oxide layer in Zr alloys consists of the monoclinic phase, which differs in its oxygen coordination numbers compared to the metastable phases. Specifically, two type of O are present, threefold and fourfold in \zro{m-}, as opposed to just fourfold O in the tetragonal and cubic phases. Thus, we expect a difference in the local atomic environment in monoclinic boundaries. In the present study, we used first-principles simulations to study the interaction of oxygen vacancies, Sn and Nb defects with the \gbsymb{} grain boundary in monoclinic \zro{}, and the effects of these defects on the structural and electronic properties of the equilibrium boundary structure and a metastable boundary structure, which is expected to be representative of a more general oxide boundary.

\section{Methodology}

\subsection{Computational details}
We performed first-principle calculations of monoclinic \zro{} grain boundaries using the CASTEP 17.2 density functional theory (DFT) code \cite{Clark2005}. The exchange-correlation functional was approximated with the Perdew, Burke and Ernzerhof (PBE) formulation of the generalized gradient approximation (GGA) \cite{Perdew1996}. Ultrasoft pseudopotentials, as generated by the CASTEP on-the-fly generator, were used for all species. A plane-wave energy cut-off of \evolt{620} and a scaling factor of the fine FFT grid of 2.75 were employed. The explicitly modelled electrons for each species, are as follows: $[\tm{O}]\,\mathrm{2s^2\,2p^4}$; $\mathrm{[\tm{Zr}]\,4s^2\,4p^6\,4d^2\,5s^2}$; $\mathrm{[\tm{Sn}]\,4d^{10}\,5s^2\,5p^2}$; $\mathrm{[\tm{Nb}]\,4p^{6}\,4d^4\,5s^1}$. In order to account for the self-interaction error present in GGA in simulations that include Nb, we utilised a Hubbard U parameter, as implemented in CASTEP \cite{Dudarev1998, Cococcioni2005}, with a value of \evolt{1.5}. This value was reported by Hautier et al.\,\cite{Hautier2012} as suitable for Nb after an extensive study on transition-metal-oxide systems that compared results from GGA+U simulations with experimental formation energies. We followed the approach of previous work in this regard \cite{Otgonbaatar2014a, Youssef2016, Bell2017}.  
   
A Monkhorst-Pack (MP) grid \cite{Monkhorst1976} with a maximum k-point spacing of \perangs{0.05} was utilised, which corresponded to k-point grids of \grid{4}{4}{4}, \grid{1}{4}{4}, \grid{1}{4}{2} and \grid{1}{2}{2} for supercells with, respectively, \grid{1}{1}{1}, \grid{6}{1}{1}, \grid{6}{1}{2} and \grid{6}{2}{2} repeats of the monoclinic \zro{} unit cell. The electronic structure properties were calculated with a denser MP grid, using a maximum spacing of \perangs{0.03}, and OptaDOS \cite{Morris2014} was used for post-processing of the DFT data to compute the density of states. We performed self-consistent-field (SCF) calculations with an energy tolerance of \evolt{1e-5}, and using the Pulay density mixing scheme \cite{Pulay1980b}. The low memory Broyden-Fletcher-Goldfarb-Shanno algorithm was used to relax ionic coordinates and, when required, supercell geometry, with energy and force convergence tolerances of \evolt{2e-5} and \evperang{0.05}, respectively. The above computational parameters were found to converge the grain boundary energies, as will be defined in Equation \ref{gbs:eqn:gbenvol}, to within \jpermsq{0.01}.

% ------------------------------------------------
\subsection{Grain boundary structures}
\begin{figure*}
    \centering

      \subcaptionbox{Supercell construction\label{gbs:fig:sups-a}}
      {{\includegraphics{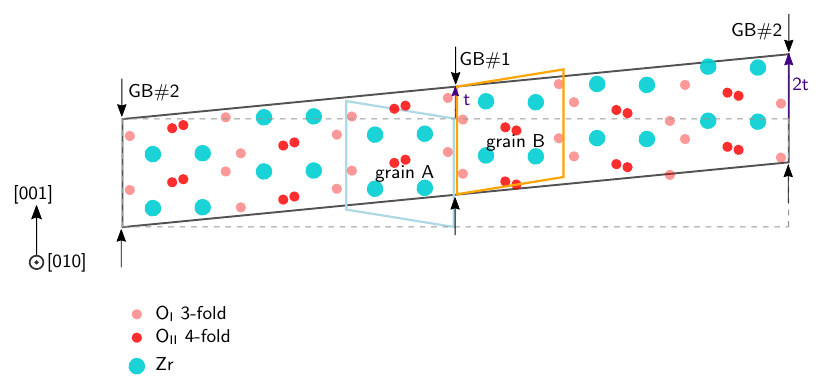}}}
      \par\bigskip
      \subcaptionbox{\gbsymb{a} (EQ)\label{gbs:fig:sups-b}}
      {{\includegraphics{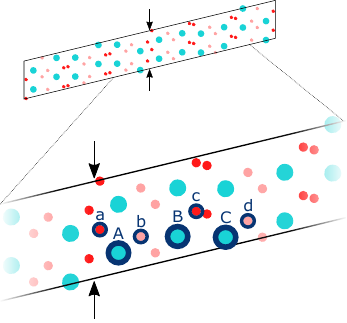}}}
      \hspace*{0.6cm}
      \subcaptionbox{\gbsymb{b} (MS)\label{gbs:fig:sups-c}}
      {{\includegraphics{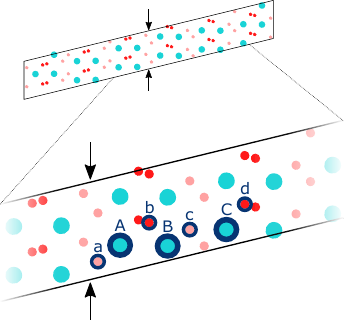}}}%
    \caption[Fully periodic supercell construction with the \gbsymb{} monoclinic \zro{} twin boundary \subref{gbs:fig:sups-a}. Unrelaxed structures of the boundary constructed with two sets of primitive crystal termination planes: \subref{gbs:fig:sups-b} \hkl(200)-\hkl(200), equilibrium structure and \subref{gbs:fig:sups-c} \hkl(100)\hkl(100), metastable structure.]{Fully periodic supercell construction with the \gbsymb{} monoclinic \zro{} twin boundary \subref{gbs:fig:sups-a}. Unrelaxed structures of the boundary constructed with two sets of primitive crystal termination planes: \subref{gbs:fig:sups-b} \hkl(200)-\hkl(200), equilibrium structure and \subref{gbs:fig:sups-c} \hkl(100)-\hkl(100), metastable structure. Denoted are also the Zr and O sites investigated as defect sites.}
    \label{gbs:fig:sups}
    \end{figure*}

% ------------------------------------------------
Starting from the experimental structure reported in \cite{Howard1988}, we firstly optimised the primitive unit cell of monoclinic \zro{}. The lattice parameters and ion positions are shown in Table \ref{gbs:tab:bu-params}, and they were found to agree well with experimental and previous simulation results \cite{Howard1988, Kuwabara2005, Jaffe2005}. Bulk \zro{m-} crystal contains two inequivalent O sites, \On{I} and \On{II}, which bond with three and four Zr atoms, respectively.    
Using the optimised bulk lattice parameters, we then constructed a supercell (see Figure \ref{gbs:fig:sups-a}) to model a twin monoclinic \zro{} boundary with a macroscopic configuration denoted by\footnote{This notation refers to: the sigma value ($\Sigma$) indicating the integer number of primitive unit cells in a coincident site lattice (CSL) used to generate a periodic supercell containing a GB; the rotation angle, the grain boundary plane, and the rotation axis. Note, monoclinic crystals have a low symmetry, and generally only a concept of a \emph{near}-CSL can be applied by straining the unit lattice. Straining the lattice was not necessary in our case, since the studied GB is a \degs{180} twin.} \gbsymb{}. For the as-constructed supercell, the grain boundary formation energy and excess volume are defined as, respectively:
  \begin{equation}
  E_{\tm{f}}^{\textrm{GB}} = \frac{E_{\textrm{tot}}^{\textrm{GB}} - E_{\textrm{tot}}^{\textrm{B}}}{2A},\qquad V_{\tm{xc}} = \frac{V^{\tm{GB}} - V^{\tm{B}}}{2A},
  \label{gbs:eqn:gbenvol}
  \end{equation}
where $E_{\textrm{tot}}^{\textrm{GB/B}}$ are the total ground state energies and $V^{\tm{GB/B}}$ are the total supercell volumes. Superscripts refer to either GB supercells or bulk \zro{m-} crystal supercells of an equivalent number of atoms. $A$ is the grain boundary area and the factor of $1/2$ accounts for the presence of two boundaries in the supercell. 

In order to find a thermodynamically plausible structure, we explored the \emph{microscopic} degrees of freedom of the boundary\,---\,the relative grain translations in the GB plane and boundary expansion, i.e.\,different excess volumes, by calculating the \gsurf{}. Since we applied full periodic boundary conditions, it was necessary to ensure that the two boundaries modelled in the supercell remain equivalent under applying any translations to the grains. We followed a method described by Guhl \cite{Guhl2015} to ensure this was the case. If we apply a translation in the GB plane, for instance $\vec{t}$, to one of the bulk grains, e.g. grain B in Figure \ref{gbs:fig:sups-a}, we must also apply a skew to the supercell of $2\vec{t}$, which preserves the inversion symmetry of both grains and, therefore, the unrelaxed boundaries remain equivalent. For each translation on the \gsurf{}, we construct structures with varying boundary expansions, and perform ion relaxation under constant volume. 
Initially, ions were relaxed only in the direction normal to the boundary in generating the full \gsurf{}. However, once a minimum energy structure was identified, full ion relaxation was performed. Convergence tests for the thickness of the supercell showed the boundary energy to vary within \jpermsq{0.01} for the chosen grain thickness of three monoclinic unit cells, compared to four unit cells.

Additionally, we examined the effect of varying the crystal termination plane on the boundary energy. We found that the lowest energy structure for the \gbsymb{} boundary is formed from \hkl(200) terminated crystals, with a translation, $\vec{t}_{\tm{min}} = (0.0, 0.5)$ \hkl[010],\hkl[001], which results in an excess volume of $V_\tm{xc}=\angs{0.03}$ and boundary energy of $E_{\tm{f}}^{\textrm{GB}} = \jpermsq{0.06}$. 

The twin boundary we are studying is a special boundary, whose properties have been suggested to be responsible for lower energy diffusion paths for species \cite{Gertsman1996, Gertsman1999, Garner2015}. Thus, we chose to compare the equilibrium (EQ) structure with a metastable (MS) local minimum structure of this boundary, which has \hkl(100) terminated grains, and a \gsurf{} minimum at $\vec{t}_{\tm{min}} = (0.0, 0.5)\hkl[010],\hkl[001]$ with $V_{\tm{xc}} = \angs{-0.08}$. 
The MS GB has a grain boundary energy of \jpermsq{0.32}, which can be considered thermodynamically feasible. For instance, boundary energies in other metal oxides, that have been reported include \jpermsq{0.6} in $\ce{HfO_2}$ \cite{McKenna2009a} and \SIrange{0.53}{1.10}{\joule\per\meter\squared} in $\ce{SrTiO_3}$ \cite{Benedek2008}. Therefore, we consider this structure as a prototype of a more general oxide boundary compared to the twin EQ GB structure. The sub-grains of the EQ GB are terminated at fourfold O sites, whereas those of the MS GB are terminated at threefold O sites. The optimised GB supercells consist of \grid{6}{1}{1} unit cell repeats and have dimensions of $b\times c$ in the boundary plane, where $b$ and $c$ are the primitive unit cell lattice parameters, and out-of-boundary sizes of \angs{31.144} and \angs{30.919} for the EQ and the MS GB, respectively.

  \begin{table}[h!]
    \caption[Lattice parameters and ion fractional coordinates of the primitive unit cell of monoclinic \zro{} as calculated in this study.]{Lattice parameters and ion fractional coordinates of the primitive unit cell of monoclinic \zro{} as calculated in this study. \On{I} and \On{II} refer to, respectively, threefold and fourfold coordinated O ions.}
    \centering
    \begin{tabular}{ll}
      \toprule
    Parameter                    & PBE DFT           \\ \hline
    $a$                             & \angs{5.175}                 \\
    $b$                             & \angs{5.248}                 \\
    $c$                             & \angs{5.358}                 \\
    $\beta$                       & $\degs{99.6}$         \\
    $(x, y, z)_{\textrm{Zr}}$     & (0.276, 0.044, 0.210) \\
    $(x, y, z)_{\textrm{O}_\tm{I}}$    & (0.068, 0.333, 0.345) \\
    $(x, y, z)_{\textrm{O}_{\tm{II}}}$ & (0.450, 0.757, 0.478) \\ \bottomrule
    \end{tabular}
    \label{gbs:tab:bu-params} 
  \end{table}
% ------------------------------------------------
\subsection{Point defect calculations}
% ------------------------------------------------
Defective bulk structures were constructed by then creating an array of \grid{l}{m}{n} copies of the relaxed bulk \zro{m-} primitive unit cell, and adding or removing one or more atoms. Defective GB structures were generated starting from the relaxed structures with \gridtwo{m}{n} unit cell repeats in the boundary plane, keeping the supercell thickness constant, and decorating only one of the boundaries in the supercell with point defect(s), as maintaining equivalence of the boundaries was not necessary at this point. The defect sites explored in this study are shown in Figures \ref{gbs:fig:sups-b} and \subref{gbs:fig:sups-c}. The defective structures were then optimised under constant volume. 

We use a modified Kr\"{o}ger-Vink notation to describe the lattice positions and electronic charges of the point defects. For a point defect, M$^q_{\tm{S}}$, M indicates the species of the defect, S corresponds to the species of the lattice site occupied by the defect, and $q$ denotes the charge state of the defect relative to the occupied site. Instead of a symbol to indicate charge as in the original notation \cite{Kroger1956}, we use a numerical digit with a plus or a minus sign, e.g. we denote $\ce{v^{..}_O}$ as \vo{2+}, which indicates an oxygen vacancy due to the removal of an $\ce{O^{''}}$ ion, leaving a net 2+ charge on the supercell. 
In this study, we considered oxygen vacancy native defects and alloying element defects (Sn and Nb), which are expected to be present on zirconium substitutional sites. For these defects, we examined the main charge states that are expected to exist under typical reactor operating conditions. Furthermore, we studied a defect cluster, consisting of an oxygen vacancy with 2+ charge and a tin substitutional defect, $\{\vo{2+} : \sn{Zr}{}\}$ with varying charge states for the combined defect. We explored three different sites for the position of \sn{Zr}{}, denoted with A, B and C in Figures \ref{gbs:fig:sups-b} and \subref{gbs:fig:sups-c} and all nearest-neighbour O sites for the position of \vo{2+}. Equivalently, a cluster consisting of \sn{Zr}{} and \vo{2+} on all nearest-neighbour O atoms in the bulk monoclinic crystal was modelled. 

In the Zhang-Northrup formalism \cite{Zhang1991}, the formation energy of a point defect, M with a charge $q$, in a grain boundary (GB) or bulk (B) system can be computed using:
  \begin{equation}
    \enX{f}{GB/B}[\defM{}] = \enX{tot}{GB/B}[\defM{}] - \enX{tot}{GB/B} + \sum_i \Delta n_i \mu_i + q \epsilon_\tm{F},
    \label{gbs:eqn:ef}
  \end{equation}

where \enX{tot}{} and \enX{tot}{}[\defM{}] are, respectively, the total energy of a pristine supercell, and that of an equivalent supercell containing a point defect \defM{}, and $\Delta n_i$ and $\mu_i$ are, respectively, the difference in the number of ions between the pristine and the defective supercells, and the chemical potential of species $i$, which has been added or removed. And finally, the Fermi level, $\epsilon_\tm{F}$, determines the chemical potential of the electrons.

In the current study, our aim is to investigate the interaction of point defects with grain boundaries, compared to a reference system consisting of bulk crystal. Therefore, the chemical potential terms in Equation \ref{gbs:eqn:ef} cancel out in any of the subsequent quantities that we compute. For conciseness in the notation, we define a defect formation energy difference:
  \begin{equation}
    \Delta \enX{f,tot}{GB/B}[\defM{}] = \enX{tot}{GB/B}[\defM{}] - \enX{tot}{GB/B}.
    \label{gbs:eqn:def}
  \end{equation}

Then, we can define a segregation energy of the point defect, which represents the energetic preference of the defect for the GB system compared to the bulk, as:
  \begin{equation}
    \enX{seg}{GB}[\defM{}] = \Delta \enX{f,tot}{GB}[\defM{}]- \Delta \enX{f,tot}{B}[\defM{}].
    \label{gbs:eqn:eseg}
  \end{equation}

The usual definition of the segregation energy uses equivalent size and shape supercells of the GB and bulk systems as a model for dilute point defect concentration. In this method, the supercells have to be large enough to minimise defect-defect interactions across periodic images and any interaction that might be present must cancel out between the two systems. It is also implicitly assumed that the physical situation we wish to model is one in which the defect remains in the dilute limit both in the bulk reservoir and in the grain boundary. We expand on this approach by exploring the dependence of the segregation energy on the defect concentrations in the GB and bulk systems. In particular, we may rewrite Equation \ref{gbs:eqn:eseg}, so that the terms are explicitly functions of defect concentrations in a GB system, \concn{}{GB}, or bulk system, \concv{}{B}, where we denote a defect number density per unit area with \concn{}{}, and defect number density per unit volume with \concv{}{}. 
  \begin{equation}
    \enX{seg}{GB}[\defM{}](\concn{}{GB}, \concv{}{B}) = \Delta \enX{f,tot}{GB}[\defM{}](\concn{}{GB})- \Delta \enX{f,tot}{B}[\defM{}](\concv{}{B}).
    \label{gbs:eqn:eseg2}
  \end{equation}

 The number of unit cell repeats, number of atoms, and volume and/or area concentrations for the corresponding bulk and GB supercell systems used in this work are listed in Table \ref{gbs:tab:conc}. We compared the effect on the segregation energy of using an equivalent size bulk supercell versus varying the bulk defect concentration. The total energy of the bulk supercell with a defect concentration in the dilute limit was extrapolated using a linear fit to the energies of supercells with \grid{2}{2}{2}, \grid{3}{3}{3} and \grid{4}{4}{4} unit cell repeats for single defects \vo{}, \sn{Zr}{} and \nb{Zr}{}, and to supercells with \grid{2}{2}{2} and \grid{3}{3}{3} repeats for defect clusters \vosn{}{}{}. The \grid{1}{1}{1} supercells were excluded from the fitting since there was clearly a very strong self-interaction effect in them.

When studying a cluster of two defects, \defM{} and \defN{}, we define a segregation energy for a second defect \defM{}, in the presence of a defect \defN{} in the grain boundary, as:

  \begin{eqnarray}
    E_{\tm{seg}}^{\tm{GB}, \defN{}}[\defM{}] = \Delta \enX{f,tot}{GB}[\defM{}, \defN{}]- \Delta \enX{f,tot}{GB}[\defN{}]\nonumber\\
    - \Delta \enX{f,tot}{B}[\defM{}].
    \label{gbs:eqn:esegcl}
  \end{eqnarray}

We can also compute a binding energy for the same defect cluster, as follows:

 \begin{eqnarray}
    \enX{bind}{GB/B}[\defM{}, \defN{}] = \Delta \enX{f,tot}{GB/B}[\defM{}, \defN{}] - \Delta \enX{f,tot}{GB/B}[\defN{}]\nonumber\\
    - \Delta \enX{f,tot}{GB/B}[\defM{}].
    \label{gbs:eqn:ebind}
  \end{eqnarray}

If the fraction of sites occupied by point defects are \fdef{GB} and \fdef{B} in the bulk and the GB system, respectively, their ratio can be expressed as \cite{McKenna2009a},
  \begin{equation}
    \frac{\fdef{GB}}{\fdef{B}} = \exp\bigg(\frac {-(\enX{f}{GB}-\enX{f}{B})} {k_\tm{B} T}\bigg) =  \exp\bigg(\frac {-\enX{seg}{GB}} {k_\tm{B} T}\bigg),
    \label{gbs:eqn:dist-def}
  \end{equation}
where $k_\tm{B}$ is the Boltzmann constant and $T$ is the temperature of the system; $T =\kelvin{603}$, an approximate reactor temperature, was considered, here. 
    \begin{table*}
      \caption[Primitive unit cell repeats, number of atoms, concentration per unit volume, $c$, and concentration per unit area, $\eta$, for the simulated supercell of each type of system.]{Primitive unit cell repeats, number of atoms, concentration per unit volume, $c$, and concentration per unit area, $\eta$, for the simulated supercell of each type of system.}
      \centering
      \begin{tabular}{@{}ccc|ccccc@{}}
        \toprule
        \multicolumn{3}{c|}{Bulk supercell}           & \multicolumn{4}{c}{Bulk/GB supercell}                                      \\ \midrule
        repeats & atoms       & \concv{}{} (\percubicnm{}) & repeats  & atoms          & \concv{}{} (\percubicnm{}) & \concn{}{} (\pernmsqr{}) & Notation\\ \midrule

      \grid{1}{1}{1}                       & 12    & 6.969                       & \grid{6}{1}{1} & 72    & 1.163                       & 3.556      & \concn{0}{}                  \\
      \grid{2}{2}{2}                       & 96    & 0.871                       & \grid{6}{1}{2} & 144   & 0.584                       & 1.778      & 0.5\concn{0}{}               \\
      \grid{3}{3}{3}                       & 324   & 0.258                       & \grid{6}{2}{1} & 144   & 0.584                       & 1.778      & 0.5\concn{0}{}               \\
      \grid{4}{4}{4}                       & 768   & 0.109                       & \grid{6}{2}{2} & 288   & 0.292                       & 0.889      & 0.25\concn{0}{}              \\
      - & $\infty$     & 0.000                           &                &       &                             &                       &       \\ \bottomrule
      \end{tabular}
      \label{gbs:tab:conc}
      \end{table*}

% ------------------------------------------------
\subsection{Structural analysis}
% ------------------------------------------------
The local atomic environment, i.e.\ the microscopic, rather than the macroscopic degrees of freedom, is an important factor in determining the properties of crystalline grain boundaries. Hence, we used several parameters to characterise the local atomic environment, and observe how they differ between the EQ and the MS GBs. The effect of introducing point defects was of particular interest.

Firstly, we calculated the volume per atom using a Voronoi tessellation, in which a volume is partitioned into convex hulls such that each hull contains one atom and every point in this convex hull is closer to that atom than any other atom. A grain boundary width was calculated based on a change in the local volume of more than 0.5\%. Next, we computed the effect of the GB on the interlayer spacing, for a stoichiometric \zro{} layer, consisting of $N$ Zr atoms, $N$ \On{I} and $N$ \On{II} atoms ($N=2$ for the smallest GB cell). If the change in the distance normal to the boundary relative to the as-constructed unrelaxed GB structure for an atom $k$ of species M is $\delta_\tm{n}(\tm{M}_k)$, the change in interlayer spacing $\Delta d_{ij}$ between layers $i$ and $j$ is defined as  

  \begin{equation}
    \Delta d_{ij} = \Delta \Bigg( \frac{\sum_{\tm{M}=(\ce{Zr}, \On{I}, \On{II})} \sum_k^N \delta_\tm{n}(\tm{M}_k)} {3N} \Bigg)_\tm{ij} - \delta_{\tm{bulk}}, 
  \end{equation}

where the first term is the average change in atom coordinates of a stoichiometric layer in the GB and $\delta_{\tm{bulk}}$ is the equivalent interlayer distance in the bulk crystal. 
We also computed the average change in distance normal to the boundary for atoms of species M, $\overbar{\delta_\tm{n}}$.

% ************************************************
\section{Results and Discussion}
% ************************************************

% ------------------------------------------------
\subsection{Pristine boundary structures}
\begin{figure*}
  \centering{
    \phantomsubcaption{\label{gbs:fig:locat-pr-a}}
    \phantomsubcaption{\label{gbs:fig:locat-pr-b}}
    \phantomsubcaption{\label{gbs:fig:locat-pr-c}}
    \phantomsubcaption{\label{gbs:fig:locat-pr-d}}
    \phantomsubcaption{\label{gbs:fig:locat-pr-e}}
    \phantomsubcaption{\label{gbs:fig:locat-pr-f}}
    {{\includegraphics[scale=1.1]{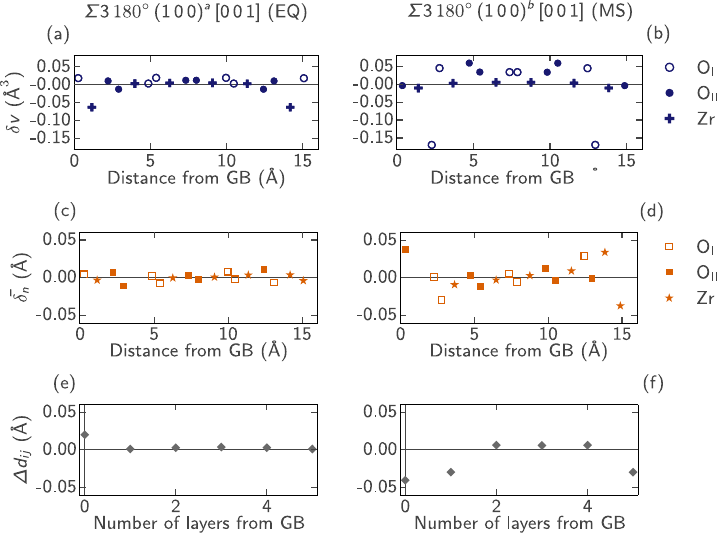}}}
  }
  \caption[Change of local atomic structural properties between the GBs and the bulk crystal: change in atomic volume, \subref{gbs:fig:locat-pr-a} and \subref{gbs:fig:locat-pr-b}; average change in atomic distance normal to the GB, \subref{gbs:fig:locat-pr-c} and \subref{gbs:fig:locat-pr-d}; change in interlayer spacing, \subref{gbs:fig:locat-pr-e} and \subref{gbs:fig:locat-pr-f}, respectively for the EQ and the MS GBs.]{Change of local atomic structural properties between the GBs and the bulk crystal. Respectively for the EQ and the MS GBs, \subref{gbs:fig:locat-pr-a} and \subref{gbs:fig:locat-pr-b}: change in atomic volume $\delta v$; \subref{gbs:fig:locat-pr-c} and \subref{gbs:fig:locat-pr-d}:  average change in atomic distance normal to the GB $\overbar{\delta_\tm{n}}$; \subref{gbs:fig:locat-pr-e} and \subref{gbs:fig:locat-pr-f}: change in interlayer spacing, $\Delta d_{ij}$.}
  \label{gbs:fig:locat-pr}
\end{figure*}

% ------------------------------------------------
Table \ref{gbs:tab:gb-props} summarises the structural properties of the two GB structures considered in this study. The EQ GB structure has a very low boundary energy of \jpermsq{0.06}, which can be attributed to the fact that it has a very bulk-like structure with no under-coordinated atoms. The GB also has a small excess volume of \angs{0.03}, which is accommodated by the first stoichiometric layer, after which the interlayer spacing converges to the bulk value, as seen in Figure \ref{gbs:fig:locat-pr-e}. The local volume per atom of the Zr atoms next to the GB decreases, whereas that of the O atoms increases. The mean of the change in distance normal to the GB is larger for the \On{II} atoms compared to the \On{I} and Zr atoms. 

\begin{table*}
  \begin{minipage}{14.6cm}
    \caption[Relaxed grain boundary expansions, widths, energies and minimum translation state on the \gsurf{} for the two grain boundary structures studied, with crystal termination planes \hkl(200)-\hkl(200), denoted by EQ (equilibrium), and \hkl(100)-\hkl(100), denoted by MS (metastable).]{Relaxed grain boundary expansions, widths, energies and minimum translation state on the \gsurf{} for the two grain boundary structures studied, with crystal termination planes \hkl(200)-\hkl(200), denoted by EQ (equilibrium), and \hkl(100)-\hkl(100), denoted by MS (metastable).}
  \centering
  \begin{tabular}{ p{2.2cm} c  c  c  c  c  c}
    \toprule	
    Grain & Crystal & Notation &  Min.&Expansion&Energy&Width\\
    boundary & termination &  &transl.&(\AA) &(Jm$^{-2}$)&(\AA) \\
    \midrule
  \raggedright $\Sigma 3$\,\degs{180} $\hkl(100)^a\,\hkl[001]$ & \hkl(200)-\hkl(200) & EQ & $(0.0, 0.5)$  & 0.03  & 0.06 & 2.3\\ \midrule
  \raggedright $\Sigma 3$\,\degs{180} $\hkl(100)^b\,\hkl[001]$ & \hkl(100)-\hkl(100) & MS & $(0.0, 0.5)$  & -0.08  & 0.32 & 4.5\\
  \bottomrule
  \end{tabular}
    \label{gbs:tab:gb-props}
    \end{minipage}
\end{table*}
The MS GB has a formation energy of \jpermsq{0.32}, which although larger than the energy of the EQ GB, is a thermodynamically plausible boundary energy for a metal oxide. As a comparison, a different monoclinic twin, the \hkl(101) tilt grain boundary in $\ce{HfO_2}$, was reported to have an energy of \jpermsq{0.60} \cite{McKenna2009a}. The MS structure has a negative excess volume of \angs{0.08} spread between the first and second layer. From the change in local atomic volume plot, we see the volume per Zr atom does not change much compared to the bulk crystal across the whole supercell. On the other hand, we observe both an expansion and contraction of the O volumes. The average change in distance is greatest at the GB, with oscillations between the Zr and O atom distances, and decays to zero after about three atomic planes. 

% ------------------------------------------------
\subsection{Segregation behaviour}
% ------------------------------------------------
\begin{figure*}
  \centering{
    \phantomsubcaption{\label{gbs:fig:eseg-d-a}}
    \phantomsubcaption{\label{gbs:fig:eseg-d-b}}
    \phantomsubcaption{\label{gbs:fig:eseg-d-c}}
    \phantomsubcaption{\label{gbs:fig:eseg-d-d}}
    \phantomsubcaption{\label{gbs:fig:eseg-d-e}}
    \phantomsubcaption{\label{gbs:fig:eseg-d-f}}
    {{\includegraphics[scale=1]{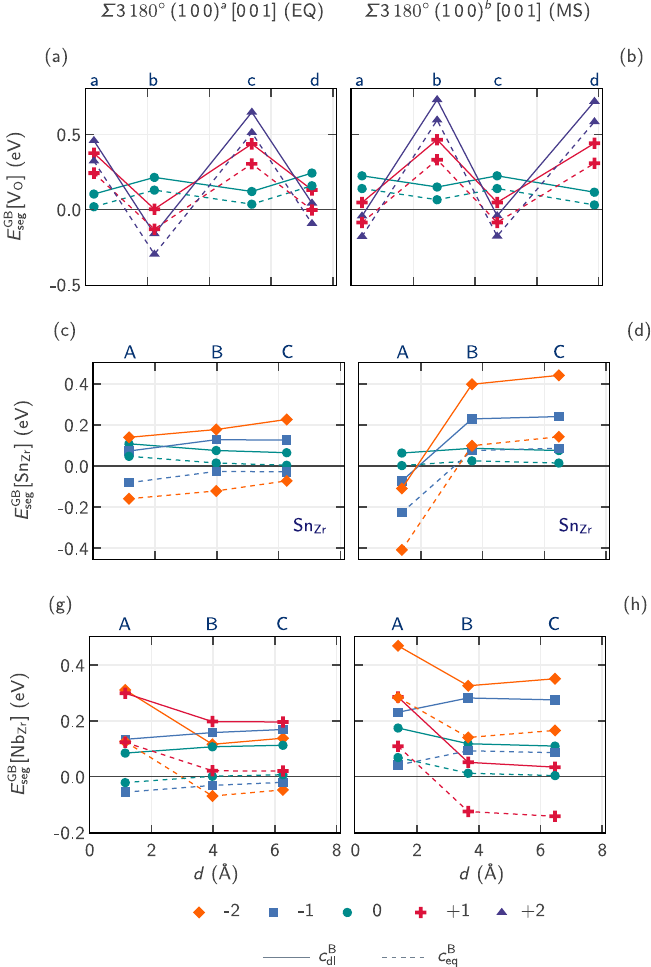}}}
  }
  \caption[Segregation energy versus distance from the GB for \ce{v_O}, \ce{Sn_{Zr}}, \ce{Nb_{Zr}} in the EQ GB: \subref{gbs:fig:eseg-d-a}, \subref{gbs:fig:eseg-d-c} and \subref{gbs:fig:eseg-d-e}; and the MS GB: \subref{gbs:fig:eseg-d-b}, \subref{gbs:fig:eseg-d-d} and \subref{gbs:fig:eseg-d-f}.]{
    Segregation energy versus distance from the GB for \ce{v_O}, \ce{Sn_{Zr}}, \ce{Nb_{Zr}} in the EQ GB: \subref{gbs:fig:eseg-d-a}, \subref{gbs:fig:eseg-d-c} and \subref{gbs:fig:eseg-d-e}; and the MS GB: \subref{gbs:fig:eseg-d-b}, \subref{gbs:fig:eseg-d-d} and \subref{gbs:fig:eseg-d-f}. Dashed lines indicate \enX{seg}{} calculated using an equivalent bulk supercell (\concv{\tm{eq}}{B}) as a reference and solid lines indicate \enX{seg}{} calculated using an extrapolated energy for a bulk crystal with defect concentration in the dilute limit (\concv{\tm{dl}}{B}). Defects sites are labelled as in Figure \ref{gbs:fig:sups}. The GB defect concentrations are 0.25\concn{0}{GB} for \vo{} and \concn{0}{GB} for \sn{Zr}{} and \nb{Zr}{}.}
  \label{gbs:fig:eseg-d}
\end{figure*}

\begin{figure*}
  \centering{
  \phantomsubcaption{\label{gbs:fig:eseg-ch-a}}
  \phantomsubcaption{\label{gbs:fig:eseg-ch-b}}
  \phantomsubcaption{\label{gbs:fig:eseg-ch-c}}
  \phantomsubcaption{\label{gbs:fig:eseg-ch-d}}
  \phantomsubcaption{\label{gbs:fig:eseg-ch-e}}
  \phantomsubcaption{\label{gbs:fig:eseg-ch-f}}
  {{\includegraphics[scale=1]{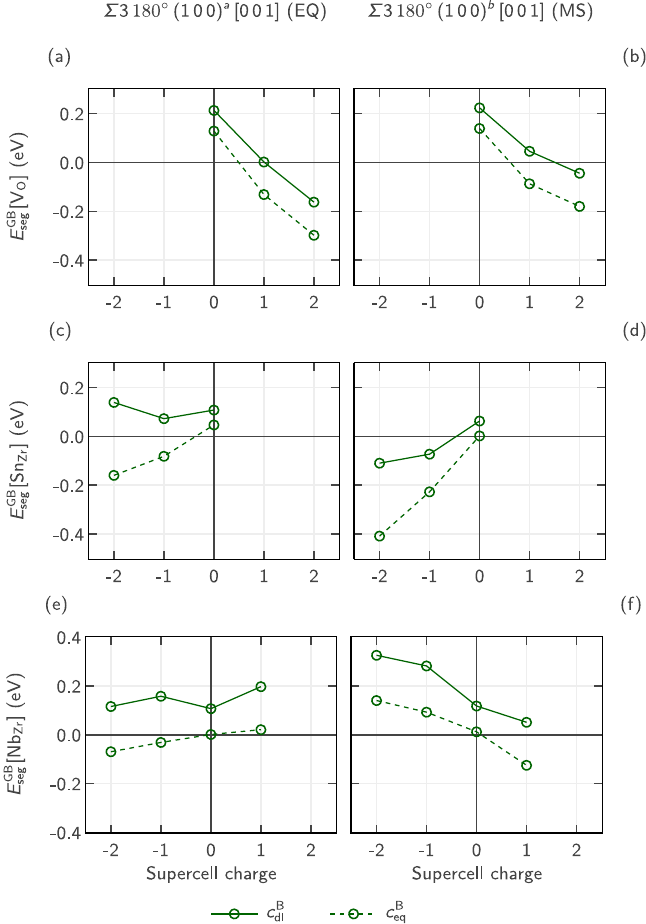}}}
  }
  \caption[Segregation energy as a function of supercell charge for the most preferred site for each defect. This represents a subset of data shown in Figure \ref{gbs:fig:eseg-d}.]{Segregation energy as a function of supercell charge for the most preferred site for each defect. This represents a subset of data shown in Figure \ref{gbs:fig:eseg-d}.}
  \label{gbs:fig:eseg-ch}
\end{figure*}
\begin{figure*}
  \centering{
  \phantomsubcaption{\label{gbs:fig:eseg-c-a}}
  \phantomsubcaption{\label{gbs:fig:eseg-c-b}}
  \phantomsubcaption{\label{gbs:fig:eseg-c-c}}
  \phantomsubcaption{\label{gbs:fig:eseg-c-d}}
  \phantomsubcaption{\label{gbs:fig:eseg-c-e}}
  \phantomsubcaption{\label{gbs:fig:eseg-c-f}}
  \phantomsubcaption{\label{gbs:fig:eseg-c-g}}
  \phantomsubcaption{\label{gbs:fig:eseg-c-h}}
  {{\includegraphics[scale=1]{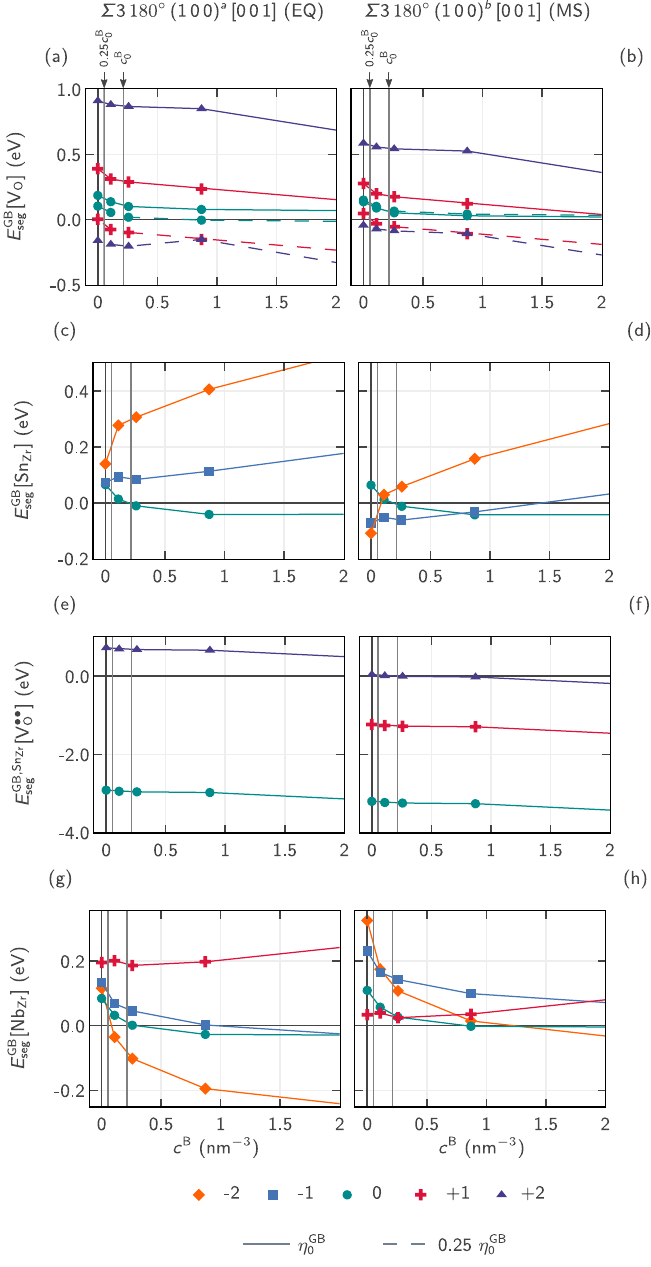}}}
  }
  \caption[Segregation energies for point defects \vo{}, \sn{Zr}{}, \vo{} in the presence of \sn{Zr}{} and \nb{Zr}{} at their most preferred sites in a given GB (as shown in Figures \ref{gbs:fig:eseg-d}) as a function of bulk defect concentration.]{
      Segregation energies for point defects \vo{}, \sn{Zr}{}, \vo{} in the presence of \sn{Zr}{} and \nb{Zr}{} at their most preferred sites in a given GB (as shown in Figures \ref{gbs:fig:eseg-d}) as a function of bulk defect concentration. Solid lines correspond to results for a GB supercell with concentration \concn{0}{GB} formed from \grid{6}{1}{1} repeats, and dashed lines apply to results for a GB supercell with concentration $0.25 \concn{0}{GB}$ formed from \grid{6}{2}{2} repeats. Vertical lines indicate the bulk concentrations, \concv{0}{B} and $0.25 \concv{0}{B}$, at which all the defects would segregate to the GB for GB concentrations, \concn{0}{GB} and $0.25 \concn{0}{GB}$, assuming a spherical grain of radius \SI{50}{\nano\meter}. 
       }
      \label{gbs:fig:eseg-c}
\end{figure*}
\begin{figure*}
  \centering{
  \phantomsubcaption{\label{gbs:fig:eseg-eta-a}}
  \phantomsubcaption{\label{gbs:fig:eseg-eta-b}}
  {{\includegraphics[scale=1]{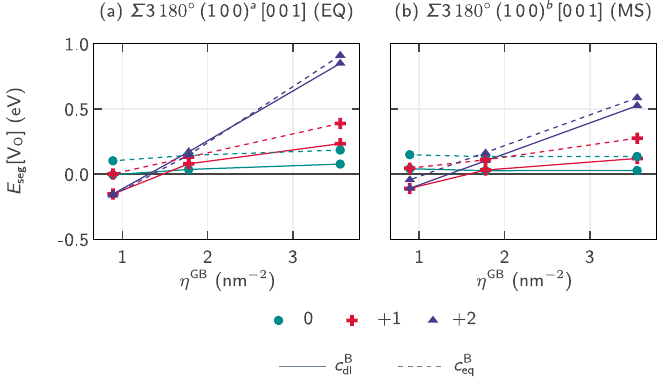}}}}
  \caption[Segregation energy for \vo{} as a function of GB concentration and supercell charge.]{
      Segregation energy for \vo{} as a function of GB concentration and supercell charge. Dashed lines indicate \enX{seg}{} calculated using an equivalent bulk supercell (\concv{\tm{eq}}{B}) as a reference and solid lines indicate \enX{seg}{} calculated using an extrapolated energy for a bulk crystal with defect concentration in the dilute limit (\concv{\tm{dl}}{B}).}
      \label{gbs:fig:eseg-eta}
\end{figure*}
\begin{figure*}
  \centering{
      \phantomsubcaption{\label{gbs:fig:c-ch-a}}
      \phantomsubcaption{\label{gbs:fig:c-ch-b}}
      \phantomsubcaption{\label{gbs:fig:c-ch-c}}
      {{\includegraphics[scale=1]{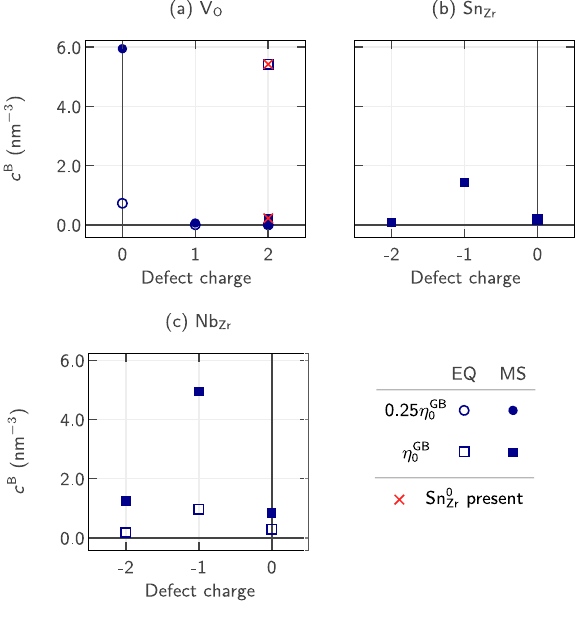}}}
  }
  \caption[Bulk concentration of \vo{}, \sn{Zr}{} and \nb{Zr}{}, at which segregation of the corresponding defect to a GB becomes favourable, as a function of charge of the defect. ]{
      Bulk concentration of \vo{}, \sn{Zr}{} and \nb{Zr}{}, at which segregation of the corresponding defect to a GB becomes favourable, as a function of charge of the defect. Charge states, for which segregation is favourable across any bulk concentration are denoted at $\concv{}{B} = 0.0$. Supercells with two grain boundary concentrations are shown with a circle and a square, for 0.25\concn{0}{GB} and \concn{0}{GB}, respectively. In the case of \vo{}, calculations for supercells containing a \sn{Zr}{0} are denoted with a red cross. 
       }
      \label{gbs:fig:c-ch}
\end{figure*}

Here, we present results on the interaction of the studied point defects with the grain boundaries and compare the segregation behaviour between the EQ and the MS grain boundary structures. The first factor we consider is the defect distance from the GB core and how it influences the segregation energy. The results are shown in Figure \ref{gbs:fig:eseg-d}. Segregation energies have been calculated using two different bulk reference systems: with a dashed line, we have plotted the results from Equation \ref{gbs:eqn:eseg}, and with a solid line, the results with reference to the chemical potential of the solute in bulk in the dilute limit. The former corresponds to the typical definition in the literature: the driving force for a grain boundary is computed as the difference in total energies between defective and pristine equivalent supercells with a GB and of bulk crystal. Based on this definition, we expect the segregation energy to converge to zero in the middle of a size-converged fully-periodic supercell, containing two GBs. An exception to this is the oscillating behaviour for \vo{} that we observe in Figures \ref{gbs:fig:eseg-d-a} and \subref{gbs:fig:eseg-d-b}. This is a manifestation of the difference in preferred bulk O site depending on charge state\,\,---\,\,a 3-fold site for charged vacancies, and a 4-fold site for neutral vacancies, as also previously demonstrated by Bell et al.\,\cite{Bell2017} and Youssef et al.\,\cite{Youssef2014}. In order to investigate how the grain boundary energetical favourability changes with charge state, we examined the effect of the supercell charge state on the segregation energy for the most preferred defect site, as shown in Figure \ref{gbs:fig:eseg-ch}. Moreover, the differences between the solid and the dashed lines in these two figures indicate the strength and nature of the self-interaction between periodic images of defects in the equivalent bulk supercell. Next, we considered how the segregation energy depends on the bulk defect concentration and, in particular for O vacancies, on the grain boundary concentration, in Figures \ref{gbs:fig:eseg-c} and \ref{gbs:fig:eseg-eta}, respectively. Vertical lines in Figure \ref{gbs:fig:eseg-c} show the bulk concentration, at which a particular grain boundary concentration would be achieved for a typical oxide grain size of \nm{50}. In other words, at what bulk concentration, for a given average grain size, all the defects would move from inside the grain to the boundary. Furthermore, in Figure \ref{gbs:fig:c-ch} we examined the bulk concentration at which segregation becomes favourable as a function of defect charge. 
\begin{figure*}
  \centering{
  \phantomsubcaption{\label{gbs:fig:eseg-bind-d-a}}
  \phantomsubcaption{\label{gbs:fig:eseg-bind-d-b}}
  \phantomsubcaption{\label{gbs:fig:eseg-bind-d-c}}
  \phantomsubcaption{\label{gbs:fig:eseg-bind-d-d}}
  {{\includegraphics[scale=1]{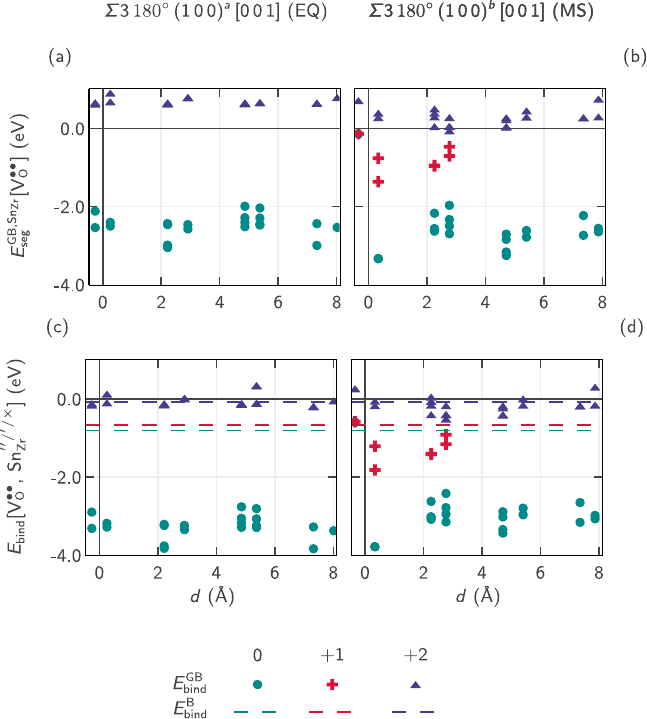}}}
  }
  \caption[Segregation energy for \vo{2+} in the presence of the \sn{Zr}{} point defect, computed with an equivalent size bulk supercell reference, \subref{gbs:fig:eseg-bind-d-a} and \subref{gbs:fig:eseg-bind-d-b}; and binding energy for defect cluster \vosn{2+}{2-/1-/0}{0/1+/2+}, \subref{gbs:fig:eseg-bind-d-c} and \subref{gbs:fig:eseg-bind-d-d}, for the EQ and MS GBs, respectively. ]{Segregation energy for \vo{2+} in the presence of the \sn{Zr}{} point defect, computed with an equivalent size bulk supercell reference, \subref{gbs:fig:eseg-bind-d-a} and \subref{gbs:fig:eseg-bind-d-b}; and binding energy for defect cluster \vosn{2+}{2-/1-/0}{0/1+/2+}, \subref{gbs:fig:eseg-bind-d-c} and \subref{gbs:fig:eseg-bind-d-d}, for the EQ and MS GBs, respectively. Charge states, denoted by colour, apply to the whole supercell. Explored sites include sites A, B and C for \sn{Zr}{}, as denoted in Figure \ref{gbs:fig:sups} and all nearest-neighbour O atoms. The average binding energy between \sn{Zr}{} and \vo{2+} positioned at all nearest-neighbour O atoms in bulk \zro{m-} is also plotted with a dashed line.}
  \label{gbs:fig:eseg-bind-d}
\end{figure*}

Figures \ref{gbs:fig:eseg-d-a} and \subref{gbs:fig:eseg-d-b} show the segregation energy for oxygen vacancies as a function of distance from the GB core, for four different O sites as indicated in the diagrams on Figures \ref{gbs:fig:sups-b} and \subref{gbs:fig:sups-c}. 
The GB concentration is $\concn{}{GB}=0.25\concn{0}{}$. We observe a segregation preference for site b in the EQ GB and for sites a and c in the MS GB for the charged defect states\,---\,all three sites are threefold O sites, which, as mentioned above, is the preferred site for charged \vo{} in bulk material. No favourability for neutral \vo{} for either GB is observed, which suggests the preference for the GBs is a result of the electrostatic interaction of the defects. This result is more clearly seen in Figures \ref{gbs:fig:eseg-ch-a} and \subref{gbs:fig:eseg-ch-b}, which show a similar trend for the two GBs of increasing favourability with increasing the magnitude of defect charge. Based on the equivalent bulk reference, the \vo{2+} defect in the EQ GB has the lowest segregation energy of \evolt{-0.3}, whereas the segregation energy in the MS GB is about \evolt{-0.2}. 
Examining Figures \ref{gbs:fig:eseg-c-a} and \subref{gbs:fig:eseg-c-b}, we note that there is no favourability for both GBs at \vo{} concentration of $\concn{0}{GB}$, whereas both GBs are favourable for charged \vo{} at GB concentration of $0.25\concn{0}{GB}$ across the majority of bulk defect concentrations, and even neutral \vo{} at some bulk concentrations. Figures \ref{gbs:fig:eseg-eta-a} and \subref{gbs:fig:eseg-eta-b} show the variation in \enX{seg}{} based on three GB supercells with concentrations as listed in Table \ref{gbs:tab:conc}. The result for $0.5\concn{0}{GB}$ was calculated as the arithmetic mean of the segregation energies for equivalent sites in the \grid{6}{1}{2} and \grid{6}{2}{1} GB supercells. The effect of GB concentration on the result for neutral vacancies is much smaller, which suggests the Coulomb interaction is the reason for the \vo{} repulsion at \concn{0}{GB}. The dependence on GB concentration appears very similar between the two GBs, but with increasing charge the favourability increase is stronger in the EQ GB than in the MS GB for the explored sites. The segregation behaviour of a charged point defect for a boundary in an ionic solid would be influenced by a combination of elastic, electrostatic and chemical interactions with the surrounding local atomic environment. As a measure of the elastic interactions, we can consider the change in local atomic volume from that in bulk for the O sites, which is shown in Figures \ref{gbs:fig:locat-pr-a} and \subref{gbs:fig:locat-pr-b}. We notice that there is a variation between positive and negative change in volume and that the explored sites are not always those with smallest volume. We can expect that a vacancy would prefer a site with negative change in volume, so there is a need to explore those sites, additionally. This might be the reason behind the slightly stronger preference for the EQ GB compared to the MS GB.

We studied the segregation energy of a Sn substitutional ion on three Zr sites in each GB structure, as shown in Figures \ref{gbs:fig:eseg-d-c} and \subref{gbs:fig:eseg-d-d}, for sites indicated in Figures \ref{gbs:fig:sups-b} and \subref{gbs:fig:sups-c}. 
Tin preferably exists as a substitutional defect on a Zr site, since \sn{}{4+} and \zr{4+} have very similar ionic radii \cite{Dean1999}. This, likely, is the reason for no GB favourability for \sn{Zr}{0} for either GB structure with reference to the bulk dilute limit. Still, when we probe the effect of increasing the bulk defect concentration, at $\concv{}{B}=\percubicnm{0.2}$, \sn{Zr}{0} experiences a bias towards both boundary structures at site A (Figures \ref{gbs:fig:eseg-c-c} and \subref{gbs:fig:eseg-c-d}). \sn{Zr}{2-}, on the other hand, exhibits the opposite behaviour. Based on \concv{\tm{eq}}{B}, the \sn{}{2+} defect is favoured at sites A and B in the EQ GB with $\eSeg{} \approx \evolt{-0.2}$ and \evolt{-0.1}, and site A in the MS GB with \evolt{-0.4}. In the dilute limit, these results are shifted, so that the EQ structure sites are no longer favourable. Accordingly, \sn{Zr}{2-} exhibits no preference at the EQ GB across the considered bulk concentrations, as demonstrated in Figure \ref{gbs:fig:eseg-c-c}, whilst it favours site A in the MS GB at very low \concv{}{B} (\ref{gbs:fig:eseg-c-d}). According to Figures \ref{gbs:fig:eseg-c-c}, \subref{gbs:fig:eseg-c-d} and \ref{gbs:fig:c-ch-b}, \sn{Zr}{1-} and \sn{Zr}{2-} display favourability only for the MS GB at \concv{}{B} less than \percubicnm{1.4} and \percubicnm{0.1}, respectively. Since the negative segregation energy decreases with decreasing charge (Figures \ref{gbs:fig:eseg-ch-c} and \subref{gbs:fig:eseg-ch-d}), the MS GB favourability is likely to be a result of electrostatic and chemical effects. 
Next, we explored the segregation energy of the doubly charged \vo{} in the presence of tin as well as the binding energy between \vo{2+} and \sn{Zr}{}, as displayed in Figures \ref{gbs:fig:eseg-bind-d-a} and \subref{gbs:fig:eseg-bind-d-b}, respectively. Calculations were performed for \concn{0}{GB} and all possible configurations for a \sn{Zr}{} at sites A, B or C and a \vo{2+} at any O site that is a nearest neighbour to either of these three sites were simulated. The charge states in the figures refer to the charge states of the combined defect cluster \vosn{2+}{2-/1-/0}{0/1+/2+}, where the oxygen vacancy always has 2+ charge and the \sn{Zr}{} can have 2+, 1+ or 0 charge. The presence of \sn{Zr}{2-} manifested most strongly in the effect on the segregation energy of \vo{2+}\,---\,values between \evolt{-2} and \evolt{-3} across all defect configurations were computed. 
This effect can be attributed to the charge neutralising effect of \sn{Zr}{2-} on \vo{2+}. For this reason, studies have suggested that \sn{Zr}{2-} increases the concentration of \vo{2+} in the oxide layer and leads to an increased corrosion rate. 
With increasing the charge of the \sn{Zr}{}, the \vo{2+} still exhibits significant segregation favourability of \SIrange{-0.2}{-1.5}{\electronvolt} in the presence of \sn{Zr}{1+}, and only borderline favourability at a few O sites in the MS GB in the presence of \sn{Zr}{0}. Figures \ref{gbs:fig:eseg-c-e} and \subref{gbs:fig:eseg-c-f} demonstrate how the bulk defect concentration impacts the segregation energy of \vo{2+} at the most favourable site near a \sn{Zr}{} defect. The strong preference for both GBs is evident across all bulk concentrations in the case of \sn{Zr}{2-}\,---\,the range of segregation energies is between \SIrange{-3.9}{-2.9}{\electronvolt} and between \SIrange{-4.2}{-3.2}{\electronvolt} for the EQ and MS GB, respectively. We note that at this GB concentration, \vo{2+} on its own did not exhibit favourability for either GB. Interestingly, \vo{2+} becomes favourable in the vicinity of the MS GB even in the presence of \sn{Zr}{0}. This result suggests that there is a chemical effect on the local atomic environment that favours \vo{2+} near the MS GB at relatively low bulk concentrations of $\concv{}{B} = \percubicnm{0.2}$, but higher concentrations are needed in the EQ GB at $\concv{}{B} = \percubicnm{5.4}$ (see Figure \ref{gbs:fig:c-ch-a}). If this chemical effect of tin on the segregation energy of doubly charged oxygen vacancies is present in other more general oxide grain boundaries, it can contribute to the reason for the reduced oxidation rate in oxides with a high fraction of twin GBs. 

Lastly, we studied the interaction of the GB structures with a \ce{Nb} substitutional defect on Zr sites. The same sites were probed as for the \sn{Zr}{} defect. The results for segregation energy as a function of distance from the GB are displayed in Figures \ref{gbs:fig:eseg-d-e} and \subref{gbs:fig:eseg-d-f} and as a function of bulk concentration in Figures \ref{gbs:fig:eseg-c-g} and \subref{gbs:fig:eseg-c-h}, for the EQ and MS GB, respectively. As expected, the Nb dilute limit in bulk reference shifts the segregation energies, and this shift has a similar magnitude across charge states for the two GB structures. It appears at a minimum for the 2- charge, followed by the zero charge, and is largest for the 1- and 1+ charge states. Based on the equivalent reference bulk, we observed a small favourability for site A in the EQ GB for charges 1- and 0, and for site B for charge state 2-. In the case of the MS GB, on the other hand, we only observe favourability for charge state 1+ at sites B and C. Figures \ref{gbs:fig:eseg-ch-e} and \subref{gbs:fig:eseg-ch-f} suggest that the difference in segregation behaviour of \nb{Zr}{} is governed by the electrostatic and chemical interactions, and that negative charge states favour the EQ GB, whereas positive charge states favour the MS GB. 
Considering next the effect of bulk defect concentration (Figures \ref{gbs:fig:eseg-c-g} and \subref{gbs:fig:eseg-c-h}), we see a strong favourability for the EQ GB of \nb{Zr}{2-} with segregation energy up to about \evolt{-0.2} across a wide range of concentrations above \percubicnm{0.18}, which is likely to be a combination of electrostatic and chemical effects since \nb{Zr}{0} exhibits similar behaviour to \sn{Zr}{0} in both GBs---that is a small preference for the interfaces at \concv{}{B} more than \percubicnm{0.2}. \nb{Zr}{1-} also favours the EQ GB, at high bulk concentrations above \percubicnm{1}, but with \eSeg{}{} only at around \evolt{-0.1}. Although, we do not see a favourability for either GB of \nb{Zr}{1+}, we might expect its segregation behaviour for the MS GB to become more favourable with the decrease of the modelled GB defect concentration, as confirmed in Figure \ref{gbs:fig:eseg-ch-f}.   
% ------------------------------------------------
\subsection{Electronic structure effects}
\begin{figure*}
  \centering{
    \phantomsubcaption{\label{gbs:fig:dos-a}}
    \phantomsubcaption{\label{gbs:fig:dos-b}}
    \phantomsubcaption{\label{gbs:fig:dos-c}}
    \phantomsubcaption{\label{gbs:fig:dos-d}}
    \phantomsubcaption{\label{gbs:fig:dos-e}}
    \phantomsubcaption{\label{gbs:fig:dos-f}}
    \phantomsubcaption{\label{gbs:fig:dos-g}}
    {{\includegraphics[scale=1]{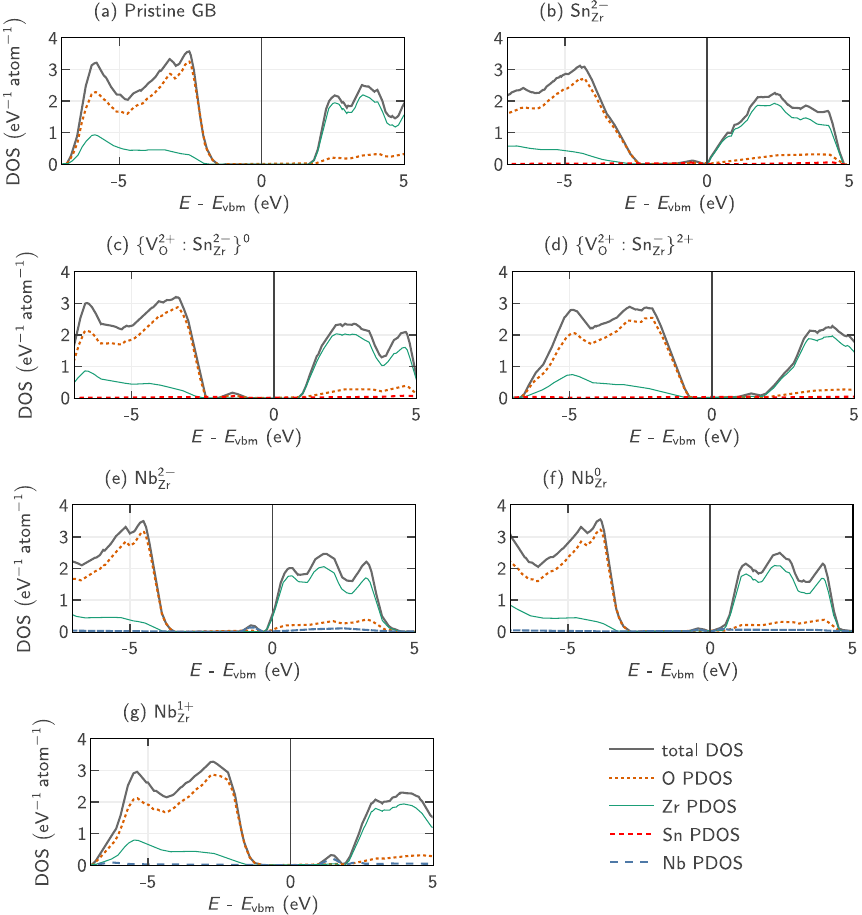}}}
      }%
  \caption[Total and projected density of states, DOS and PDOS, respectively, in a pristine \gbsymb{} GB \subref{gbs:fig:dos-a}; with a \sn{Zr}{2-} defect at site A \subref{gbs:fig:dos-b}; with a \vosn{2+}{2-}{0} defect cluster \subref{gbs:fig:dos-c}, with a \vosn{2+}{0}{2+} defect cluster \subref{gbs:fig:dos-d}; with a \nb{Zr}{2-} defect at site B \subref{gbs:fig:dos-e}; with a \nb{Zr}{0} defect at site B \subref{gbs:fig:dos-f} and with a \nb{Zr}{1+} defect at site B \subref{gbs:fig:dos-g}.]{Total and projected density of states, DOS and PDOS, respectively, in a pristine \gbsymb{} GB \subref{gbs:fig:dos-a}; with a \sn{Zr}{2-} defect at site A \subref{gbs:fig:dos-b}; with a \vosn{2+}{2-}{0} defect cluster \subref{gbs:fig:dos-c}, with a \vosn{2+}{0}{2+} defect cluster \subref{gbs:fig:dos-d}; with a \nb{Zr}{2-} defect at site B \subref{gbs:fig:dos-e}; with a \nb{Zr}{0} defect at site B \subref{gbs:fig:dos-f} and with a \nb{Zr}{1+} defect at site B \subref{gbs:fig:dos-g}. The Fermi level is indicated with reference to the valence band maximum (\enX{vbm}{}). No difference was observed in the DOS between the EQ and MS GBs, so the EQ structure results were plotted.
  }\label{gbs:fig:dos}
\end{figure*}

% ------------------------------------------------
Figure \ref{gbs:fig:dos} demonstrates how the total electron density of states (DOS) of the \zro{} boundary changes as a result of the introduction of point defects. Also, shown are the projected density of states (PDOS) onto the Zr and O atoms, and any Sn or Nb atoms present. The results for both the EQ and the MS GBs were very similar, so here we show the DOS of the equilibrium structure. Firstly, we see no change in the DOS of the pristine boundary compared to a bulk unit cell, which would be expected as the local atomic environment near the boundary is very bulk-like with no undercoordinated atoms (Figure \ref{gbs:fig:dos-a}). Another expected result is the underestimation of the experimental band gap by the PBE exchange-correlation functional. Using a hybrid functional, such as \cite{Becke1993}, would correct the band gap, however, the computational expense for the number and size of the performed calculations was unfeasible. Furthermore, this was not necessary, as our aim is to observe the relative effect of dopant gap states on the band gap.

Figure \ref{gbs:fig:dos-b} demonstrates the effect on the DOS of the presence of \sn{Zr}{2-} at the most preferred site A in the GBs. Similarly as in bulk, an occupied gap state emerges, which was not present in the case of the presence of \sn{Zr}{0}. In Figures \ref{gbs:fig:dos-c} and \subref{gbs:fig:dos-d}, we note the emergence of an occupied and an unoccupied state, in the presence of defect clusters \vosn{2+}{2-}{0} and \vosn{2+}{0}{2+}, respectively. Closer inspection, allows us to see that in the former, there are contributions from the O and the Sn PDOS, whereas in the latter the contributions are from all three species. Furthermore, the emergent gap state in the  \vosn{2+}{0}{2+} cluster was not observed in the bulk phase.  Figures \ref{gbs:fig:dos-e}, \subref{gbs:fig:dos-f} and \subref{gbs:fig:dos-g} show us that for \nb{Zr}{2-} and \nb{Zr}{0}, filled electronic states appear in the band gap, whereas for the \nb{Zr}{1+} an unfilled state appears. The same electronic states for these defects were observed in the bulk tetragonal \zro{} phase by Otgonbaatar et al.\,\cite{Otgonbaatar2014a} and we observed in the bulk monoclinic \zro{} phase.

% ------------------------------------------------
\subsection{Implications for Zr oxidation}
% ------------------------------------------------
As mentioned in the introduction, studies have demonstrated that in a slow moving protective oxide, a gradient of oxygen across the interfaces develops---from O in solid solution in the metal, through the formation of a sub-oxide \ce{ZrO} phase \cite{Gabory2015a, Nicholls2014}, to the stochiometric \zro{} phases. Ma et al.\, recalculated the Zr-O phase diagram to include metallographic and X-ray experimental data that showed that \zro{m-} and \zro{t-} can exist as non-stochiometric compounds. Therefore, we can assume that the oxygen vacancy concentration in the \zro{} protective oxide layer next to the metal-oxide interface, would be at least 0.5\,at\%, equal to \percubicnm{0.42}, or higher than that in slow growing oxides with the presence of the tetragonal phase, which is known to be stabilised by oxygen vacancies. Based on our results, the vacancy concentration with 2+ charge in the studied boundaries would be at least \pernmsqr{0.89} (Figure \ref{gbs:fig:eseg-c-a},\,\subref{gbs:fig:eseg-c-b}) at these bulk concentrations. We do not find any preference for neutral vacancies, however. If we consider a range of Sn dopant concentrations between 0.1\,at\% and 1.0\,at\% in the metal, corresponding to between \percubicnm{0.03} and \percubicnm{0.28} as substitutional defects in the oxide, and if we assume a \nm{50} spherical grain, any present \ce{Sn^4+} would marginally prefer both GB structures and this preference will increase with increasing Sn concentration. Conversely, we expect stronger preference of the lower valence states for the MS GB, and consequently more general GBs, compared to the twin EQ structure. Combining this observation, with the effect that Sn has on the \vo{2+} concentration, we see a significant increase of vacancy segregation to the MS GB. This result would suggest an adverse effect of Sn on the oxidation behaviour. Of course, the opposing view is a potential positive effect due to binding of Sn with \vo{2+}, which was found to be significantly stronger in both of the GB structures than in the bulk lattice. However, we found the binding energies to be very similar in both boundaries, whereas the \vo{2+} segregation energy to the MS GB decreased by about \evolt{3} and \evolt{0.5} in the presence of \sn{}{2+} and \sn{}{4+}, respectively, bringing the absolute segregation energies to about \evolt{-3} and \evolt{0.05}. Therefore, this effect might make \vo{} transport harder through the low energy twin studied here compared to more general boundaries.

\begin{table*}
  \caption[Fraction of occupied O sites and separation between \vo{2+} in bulk crystal and the EQ GB for a range of fractions of bulk sites occupied by vacancies.]
  {Fraction of occupied O sites and separation between \vo{2+} in bulk crystal and the EQ GB for a range of fractions of bulk sites occupied by vacancies calculated using Equations \ref{gbs:eqn:dist-def} and based on \enX{seg}{GB}[\vo{2+}](\concn{}{GB}=0.25\concn{0}{GB}, \concv{}{B}=0.0). The corresponding volume and area concentrations are also computed.}
  \centering
  \begin{tabular}{@{}cccccc@{}}
    \toprule
    \fdef{B} (\%) & \concv{}{B} (\percubicnm{}) & \fdef{GB} (\%) & \concn{}{GB} (\pernmsqr{}) & \ddef{B} (\angs{}) & \ddef{GB} (\angs{}) \\ \midrule
    0.001         & 0.001                       & 0.02           & 0.003                      & 121.5              & 176.0               \\
    0.01          & 0.01                        & 0.23           & 0.03                       & 56.4               & 55.7                \\
    0.1           & 0.06                        & 2.27           & 0.32                       & 26.2               & 17.6                \\
    0.3           & 0.17                        & 6.81           & 0.97                       & 18.1               & 10.2                \\ \bottomrule
    \end{tabular}
  \label{gbs:tab:def-dist}
\end{table*}

Nb is added in various concentrations of up to 2.5\,at\% \cite{Motta2015}, corresponding to \percubicnm{0.70}. As seen in Figure \ref{gbs:fig:eseg-c-g}, \nb{Zr}{2-} has the opposite effect to \sn{Zr}{2-}, and increasingly favours the EQ GB with a segregation energy up to about \evolt{-0.2}, and we might expect it to bind with \vo{2+}, similarly to the case in the tetragonal phase \cite{Bell2017}. It would be interesting to see its effect on the vacancies' segregation energy, which is planned in future work. In the MS GB, if \nb{Zr}{1+} is favourable, it might balance the negative charge from the \sn{Zr}{2-} defect, and therefore, reduce its detrimental effect. 
Similarly, it was also previously suggested that \nb{Zr}{1+} might suppress the formation of \vo{2+}, so reducing ion conductivity \cite{Bell2015}. 

Since the electron transport in the oxide layer is hypothesised to be the corrosion rate limiting factor, as mentioned earlier, it is important to examine what effect point defects might have. One proposed mechanism for the electron transport in an insulator such as zirconia involves the electron hopping through emerging gap states in the band gap. McKenna et al.\,\cite{McKenna2009a} found evidence for this transport mechanism in a monoclinic \ce{HfO_2} twin GB through neutral oxygen vacancies segregating to the boundary and resulting in gap states, eventually closing the band gap at high enough concentrations. We found that neutral oxygen vacancies in the studied \zro{} GB also result in a filled gap state in the band gap, but we did not find them favourable near the GBs. However, positive oxygen vacancies are favourable and in the presence of \sn{Zr}{2-}, a gap state emerges, as seen in Figure \ref{gbs:fig:dos-c}. Moreover, in the presence of of \sn{Zr}{0}, an unoccupied gap state, which is only present in the boundary and not in the bulk phase appeared. If we consider a range of fractions of O sites occupied by \vo{2+} in the bulk crystal, and the respective volume concentrations, we can calculate the corresponding fractions of sites and area concentrations in the GB, as well as the typical separation distance between any two vacancies, as shown in Table \ref{gbs:tab:def-dist}. We see that for a wide range of bulk concentrations, the vacancies are closer in the GB compared to in the bulk. Electron tunnelling rates depend exponentially on the separation between defects \cite{McKenna2009a}, which suggests boundaries could act as percolation paths for electrons. This effect might be present even in the absence of tin, if electrons happen to sit on the oxygen vacancy and neutralise it; a gap state would be created. In the presence of tin in the boundaries, it seems the effect would be exacerbated, especially, in the MS GB, where we observed favourability for Sn. Nb defects might also take part in the electron transport, as they too introduce gap states in the band gap. It would be interesting to examine further the effect of clusters containing all three defects on the structural and electronic properties of the boundaries.  
% ************************************************
\section{Conclusions}
% ************************************************
In this study, we modelled a \gbsymb{} twin grain boundary in monoclinic \zro{} using density functional theory. We studied the structural and electronic properties and the effects of introducing point defects to the equilibrium (EQ) structure of the boundary and a metastable structure (MS), which was used as a representation of a more general monoclinic GB. The findings of our study can be summarised as follows:
\begin{itemize}
  \item Across a wide range of bulk defect concentrations, we found that \vo{2+} preferentially segregates to both interfaces.
  \item \sn{Zr}{0} shows preference for both boundaries, whereas \sn{Zr}{2-} shows only preference for the MS GB.
  \item A stronger binding near both GB structures compared to that in bulk for the \vosn{2+}{2-}{0} defect cluster was found.  
  \item In the presence of \sn{Zr}{2-} and \sn{Zr}{0}, the favourability of oxygen vacancies for the MS GB increases to a value of about \evolt{-3} and \evolt{-0.1}, respectively. This suggests Sn might lead to an increase of oxygen vacancies near higher energy boundaries, and therefore, an increase in oxidation. 
  \item \nb{Zr}{2-} favours the EQ GB across a wide range of bulk concentrations with segregation energy of \evolt{-0.2}, whereas \nb{Zr}{1+} favours the MS GB under certain concentrations. The latter might interact with \sn{Zr}{2-} and reduce additional oxygen vacancies usually compensating the tin defect. 
  \item \sn{Zr}{2-}, defect clusters \vosn{2+}{2-}{0}, as well as \nb{Zr}{} with oxidation states from 2- to 0 introduce a gap state in the band gap, which reduces the band gap. For the defect cluster, the effect is only present in a boundary and not in the bulk phase.
  \item Monoclinic GBs might act as percolation paths for electron transport, an effect expected to be increased in the presence of Sn.
  \item The equilibrium structure of \gbsymb{} in monoclinic \zro{} was found to have very low boundary energy, with very similar atomic structure to the bulk crystal, and to favour \vo{2+} and \nb{Zr}{2-} but not \sn{Zr}{2-}, which might contribute to making this interface a difficult diffusion path for species. 

\end{itemize}

% ************************************************
\section*{Acknowledgements}
% ************************************************
The authors gratefully acknowledge funding from the Engineering and Physical Sciences Research Council UK (EPSRC) through the Centre for Doctoral Training in Advanced Metallic Systems. Computational resources at the Computational Shared Facility at the University of Manchester and assistance provided by the Research IT team are also acknowledged. The authors would like to thank Adam J.\,Plowman for assistance with simulation management. C.\,P.\,Race was funded by a University Research Fellowship of the Royal Society.

\bibliography{bibliography}

\end{document}

%% file: commands.tex
\newcommand{\overbar}[1]{\mkern 3mu\overline{\mkern-3mu#1\mkern-3mu}\mkern 3mu}
\renewcommand{\vec}[1]{\mathbf{#1}}
\newcommand{\tm}{\mathrm}

\newcommand{\jpermsq}[1]{\SI[]{#1}{\joule\per\meter\squared}}
\newcommand{\evperang}[1]{\SI[]{#1}{\electronvolt\per\angstrom}}
\newcommand{\evolt}[1]{\SI[]{#1}{\electronvolt}}
\newcommand{\angs}[1]{\SI[]{#1}{\angstrom}}
\newcommand{\perangs}[1]{\SI[]{#1}{\per\angstrom}}
\newcommand{\degs}[1]{\SI[]{#1}{\degree}}
\newcommand{\kelvin}[1]{\SI{0}{\kelvin}}
\newcommand{\celsius}[1]{\SI{#1}{\degreeCelsius}}
\newcommand{\metre}[1]{\SI{#1}{\metre}}
\newcommand{\nm}[1]{\SI{#1}{\nano\metre}}
\newcommand{\percubicnm}[1]{\SI{#1}{\per\cubic\nano\metre}}
\newcommand{\pernmsqr}[1]{\SI{#1}{\per\square\nano\metre}}

\newcommand{\nb}[2]{\ensuremath{\ce{Nb^{#2}_{#1}}}}

\newcommand{\On}[1]{\ce{O_{#1}}}
\newcommand{\sn}[2]{\ensuremath{\ce{Sn^{#2}_{#1}}}}
\newcommand{\vo}[1]{\ensuremath{\ce{V^{#1}_O}}}
\newcommand{\vosn}[3]{\ensuremath{\{\ce{V^{#1}_O}:\ce{Sn^{#2}_{Zr}}\}\ce{^{#3}}}}
\newcommand{\zr}[1]{\ensuremath{\ce{Zr}^{#1}}}
\newcommand{\zro}[1]{{#1}\ce{ZrO_2}}

\newcommand{\concv}[2]{\ensuremath{c_{#1}^\tm{#2}}}
\newcommand{\concn}[2]{\ensuremath{\eta_{#1}^\tm{#2}}}
\newcommand{\ddef}[1]{\ensuremath{d^\tm{#1}}}
\newcommand{\defM}{\ensuremath{\tm{M}^q}}
\newcommand{\defN}{\ensuremath{\tm{N}^p}}
\newcommand{\eSeg}[1]{\ensuremath{E_\tm{seg}^\tm{#1}}}
\newcommand{\enX}[2]{\ensuremath{E_\tm{#1}^\tm{#2}}}
\newcommand{\fdef}[1]{\ensuremath{f^\tm{#1}}}
\newcommand{\gbsymb}[1]{\ensuremath{\Sigma3}\,\ensuremath{\degs{180}}\,\ensuremath{\hkl(100)^{#1}}\,\hkl[001]}
\newcommand{\grid}[3]{\ensuremath{#1\times#2\times#3}}
\newcommand{\gridtwo}[2]{\ensuremath{#1\times#2}}
\newcommand{\gsurf}{\ensuremath{\gamma} surface}